\documentclass[10pt]{article}
\usepackage{graphicx}
\usepackage{psfig}
\begin{document}

\newcommand{\beq}{\begin{equation}}
\newcommand{\eeq}[1]{\label{#1} \end{equation}}
\newcommand{\insertplot}[1]{\centerline{\psfig{figure={#1},width=15.0cm}}}

\parskip=0.3cm

\begin{titlepage}

\vskip 0.9cm

\centerline{\Large \bf On effects of non-Euclidean geometry in
quantum theory}

\vskip 0.3cm

\centerline{Yu.A. Sitenko $ ^{a\dagger}$, N.D. Vlasii
$^{ab\star}$}

\vskip 0.1cm

\centerline{$^{a}$
 \sl  Bogolyubov Institute for Theoretical
Physics, National Academy of Sciences of Ukraine,} \centerline{\sl
14 b Metrologichna str., Kyiv 03143 , Ukraine}

\centerline{$^{b}$ \sl Department of Physics, Taras Shevchenko
National University of Kyiv,}  \centerline{\sl 6 Academician
Glushkov ave., Kyiv 03680, Ukraine }

\vskip 0.2cm

\begin{abstract}
Theory of scattering of a quantum-mechanical particle on a cosmic
string is developed. S-matrix and scattering amplitude are
determined as functions of the flux and the tension of the string.
We reveal that, in the case of the nonvanishing tension, the
high-frequency limit of the differential scattering cross section
does not coincide with the differential cross section for
scattering of a classical pointlike particle on a string.
\end{abstract}

\vskip 0.1cm


\vskip 0.1cm

(based on talks given at: the International Workshop "Frontiers of
Particle Astrophysics", June 21-24, 2004, Kyiv, Ukraine; the
George Gamow Memorial International Conference "Astrophysics and
Cosmology after Gamow - Theory and Observations", August 8-14,
2004, Odessa, Ukraine; the IV International Conference
"Non-Euclidean Geometry in Modern Physics and Mathematics",
September 7-11, 2004, Nizhny Novgorod, Russia)

\vskip 10cm


$
\begin{array}{ll}
^{\dagger}\mbox{{\it e-mail address:}} &
   \mbox{yusitenko@bitp.kiev.ua} \\
^{\star}\mbox{{\it e-mail address:}} & \mbox{vlasii@bitp.kiev.ua}
\end{array}
$

\end{titlepage}

\section{Introduction}
\label{s1}

Usually, the effects of non-Euclidean geometry are identified with
the effects which are due to the curvature of space. It can be
immediately shown that this is not the case and there are spaces
which are flat but non-Euclidean.

A simplest example is given by a twodimensional space (surface)
which is obtained from a plane by cutting a segment of a certain
angular size and then sewing together the edges. The resulting
surface is the conical surface which is flat but has a singular
point corresponding to the apex of the cone. To be more precise,
the intrinsic (Gauss) curvature of the conical surface is
proportional to the twodimensional delta-function placed at the
apex; the coefficient of proportionality is the deficit angle.
Usual cones correspond to positive values of the deficit angle,
i.e. to the situation when a segment is deleted from the plane.
But one can imagine a situation when a segment is added to the
plane; then the deficit angle is negative, and the resulting flat
surface can be denoted as a saddle-like cone. The deleted segment
is bounded by the value of $2\pi$, whereas the added segment is
unbounded. Thus, deficit angles for possible conical surfaces
range from $-\infty$ to $2\pi$.

It is evident that an apex of the conical surface with the
positive deficit angle can play a role of the convex lens, whereas
an apex of the conical surface with the negative deficit angle can
play a role of the concave lens. Really, two parallel trajectories
coming from infinity towards the apex from different sides of it,
after bypassing it, converge (and intersect) in the case of the
positive deficit angle, and diverge in the case of the negative
deficit angle. This demonstrates the non-Euclidean nature of
conical surfaces. It is interesting that this item provides a
basis for understanding such physical objects as cosmic strings.
In the present paper we shall discuss peculiarities of quantum
theory and its quasiclassical limit, which are due to
non-Euclidean geometry of locally flat space-times.

\section {Space-time in the presence of a cosmic string}\label{concl}

Cosmic strings are topological defects which are formed as a
result of phase transitions with spontaneous breakdown of
symmetries at early stages of evolution of the universe, see,
e.g., reviews in Refs.\cite{Hi,Vi}. In general, a cosmic string is
characterized by two quantities: \noindent flux
\begin{equation}
\Phi=\int\limits_{\rm core}d^2x\sqrt{g}B^3,
\end{equation}
\noindent and tension
\begin{equation}
\mu=\frac{1}{16\pi G}\int\limits_{\rm core}d^2x\sqrt{g}R;
\end{equation}
here the integration is over the transverse section of the core of
the string, $B^3$ is the field strength which is directed along
the string axis, $R$ is the scalar curvature, $G$ is the
gravitational constant, and units $\hbar=c=1$ are used. The
space-time metric outside the string core is
\begin{equation}
ds^2=dt^2-(1-4G\mu)^{-1}d{\stackrel{\sim}{r}}^2-
(1-4G\mu)\stackrel{\sim}{r}^2d\varphi^2-dz^2=
dt^2-dr^2-r^2d\stackrel{\sim}{\varphi}^2-dz^2,
\end{equation}
where
\begin{equation}
\stackrel{\sim}{r}=r\sqrt{1-4G\mu}, \quad 0\leq\varphi<2\pi,\quad
0\leq\stackrel{\sim}{\varphi}<2\pi (1-4G\mu).
\end{equation}
A surface which is transverse to the axis of the string is
isometric to the surface of a cone with a deficit angle equal to
$8\pi G\mu$. Such space-times were known a long time ago
(M.~Fierz, unpublished, see footnote in Ref.\cite{We}) and were
studied in detail by Marder \cite{Ma}. In the present context, as
cosmological objects and under the name of cosmic strings, they
were introduced in seminal works of Kibble \cite{Ki} and Vilenkin
\cite{Vil}. A cosmic string resulting from a phase transition at
the scale of the grand unification of all interactions is
characterized by the values of tension
\begin{equation}
\mu\sim(10^{-7}\div10^{-6})G^{-1}.
\end{equation}
The nonvanishing of the string tension leads to various
cosmological consequences and, among them, to a very distinctive
gravitational lensing effect. A possible observation of such an
effect has been reported recently \cite{Sa}, and this has revived
an interest towards cosmic strings.

The flux parameter (1) is nonvanishing for the so-called gauge
cosmic strings, i.e. strings corresponding to spontaneous
breakdown of local symmetries. If tension vanishes ($\mu=0$), then
a gauge cosmic string becomes a magnetic string, i.e. a tube of
the magnetic flux lines in Euclidean space. If the tube is
impenetrable for quantum-mechanical charged particles, then
scattering of the latter on the magnetic string depends on flux
$\Phi$ periodically with period $2\pi e^{-1}$ ($e$ is the coupling
constant~-- charge of the particle). This is known as the
Bohm-Aharonov effect \cite{Aha}, which has no analogue in
classical physics, since the classical motion of charged particles
cannot be affected by the magnetic flux from the impenetrable for
the particles region. The natural question is, how the
nonvanishing string tension ($\mu\neq0$) influences scattering of
quantum-mechanical particles on the string. Thus, the subject of
cosmic strings, in addition to tantalizing phenomenological
applications, acquires a certain conceptual importance.

\section {Quantum scattering on a cosmic string}\label{concl}

Due to nonvanishing flux $\Phi$ and tension $\mu$, the quantum
scattering of a test particle by a cosmic string is a highly
nontrivial problem. It is impossible to choose a plane wave as the
incident wave, because of the long-range nature of the interaction
inherent in this problem. A general approach to quantum scattering
in the case of long-range interactions was elaborated by Hormander
\cite{Ho}. This approach covers the cases of scattering on a
Coulomb center and on a magnetic string ($\mu=0$), but is not
applicable to the case of scattering on a cosmic string ($\mu\neq
0$). Therefore the last case needs a special consideration and a
thorough substantiation.

When the effects of the core structure of a cosmic string are
neglected and the transverse size of the core is negligible, the
field strength and the scalar curvature are presented by
twodimensional delta-functions. Scattering of a quantum-mechanical
particle on an idealized (without structure) cosmic string was
considered in Refs.\cite{Hoo,De,So,Si2}. A general theory of
quantum-mechanical scattering on a cosmic string, permitting to
take into account the effects of the core structure, was
elaborated in Ref.\cite{Si5}. According to this theory, the
$S$-matrix in the momentum representation is
\begin{eqnarray}
S(k,\varphi;\,\,k',\varphi')=\frac{1}{2}\frac{\delta(k-k')}{\sqrt{kk'}}\left\{\Delta
(\varphi-\varphi'+\frac{4G\mu\pi}{1-4G\mu})\exp\left[-\frac{ie\Phi}{2(1-4G\mu)}
\right]+\right. \nonumber \\
+\left.\Delta\left(\varphi-\varphi'-\frac{4G\mu\pi}{1-4G\mu}\right)
\exp\left[\frac{ie\Phi}{2(1-4G\mu)}\right]\right\}+\delta(k-k')
\sqrt{\frac{i}{2\pi k}}f(k,\,\,\varphi-\varphi'),
\end{eqnarray}
where the initial $({\bf k})$ and final $({\bf k}')$
twodimensional momenta of the particle are written in polar
variables, $f(k,\,\,\varphi-\varphi')$ is the scattering
amplitude, and $\Delta(\varphi)=
\frac{1}{2\pi}\sum\limits_{n=-\infty}^{\infty}e^{in\varphi}$ is
the angular part of the twodimensional delta-function. Note that
in the case of short-range interaction one has
$2\Delta(\varphi-\varphi')$ instead of the figure brackets in
Eq.(6). Thus, one can see that, due to the long-range nature of
interaction, even the conventional relation between $S$-matrix and
scattering amplitude is changed, involving now a distorted unity
matrix (first term in Eq.(6)) instead of the usual one,
$\delta(k-k')\Delta(\varphi-\varphi')(kk')^{-1/2}$.

In view of the comparison with the Bohm-Aharonov effect
\cite{Aha}, we shall be interested in the situation when the
string core is impenetrable for the particle. The scattering
amplitude in this case takes form:
\begin{equation}
f(k,\,\,\varphi)=f_0(k,\,\,\varphi)-\sqrt{\frac{2}{\pi
ik}}\sum\limits_{n=-\infty}^{\infty}\exp[in\varphi-i(\alpha_n-|n|)\pi]
\frac{J_{\alpha_n}(kr_c)}{H_{\alpha_n}^{(1)}(kr_c)},
\end{equation}
where $r_c$ is the radius of the string core, $J_\nu(u)$ and
$H_\nu^{(1)}(u)$ are the Bessel and the first-kind Hankel
functions of order $\nu$,
\begin{equation}
\alpha_n=\left|n-\frac{e\Phi}{2\pi}\right|(1-4G\mu)^{-1},
\end{equation}
and
\begin{eqnarray}
f_0(k,\varphi)&=&\frac{1}{\sqrt{2\pi i k}} \left\{
\frac{\exp\left[i[[\frac{e\Phi}{2\pi}]](\varphi+
\frac{4G\mu\pi}{1-4G_\mu})-\frac{ie\Phi}{2(1-4G\mu)}\right]}{1-\exp
\left[-i\left(\varphi+\frac{4G\mu\pi}{1-4G\mu}\right)\right]}-\right.
\nonumber \\
&-&\left.\frac{\exp\left[i[[\frac{e\Phi}{2\pi}]]\left(\varphi-
\frac{4G\mu\pi}{1-4G\mu}\right)+\frac{ie\Phi}{2(1-4G\mu)}\right]}{1
-\exp\left[-i\left(\varphi-\frac{4G\mu\pi}{1-4G\mu}\right)\right]}
\right\}
\end{eqnarray}
is the amplitude of scattering on an idealized (without structure)
cosmic string, $[[u]]$ is the integer part of $u$. Sum over $n$ in
Eq.(7) describes the core structure effects. In the low-frequency
limit $(k\rightarrow 0)$ these effects die out, and the
differential cross section (i. e. the square of the absolute value
of the amplitude) takes form
\begin{eqnarray}
\frac{d\sigma}{d\varphi}&=&\frac{1}{4\pi k} \left\{
\frac{1}{2\sin^2\left[\frac{1}{2}\left(\varphi+\frac{4G\mu\pi}{1-
4G\mu}\right)\right]}+\frac{1}{2\sin^2\left[\frac{1}{2}\left(\varphi-
\frac{4G\mu\pi}{1-4G\mu}\right)\right]}-\right. \nonumber \\
&-&\left.
\frac{\cos\left[\frac{e\Phi}{1-4G\mu}-\left(2[[\frac{e\Phi}{2\pi}]]+1\right)
\frac{4G\mu\pi}{1-4G\mu}\right]}{\sin\left[\frac{1}{2}\left(\varphi+
\frac{4G\mu\pi}{1-4G\mu}\right)\right]\sin\left[\frac{1}{2}\left(\varphi-
\frac{4G\mu\pi}{1-4G\mu}\right)\right]} \right\}.
\end{eqnarray}

\section {Differential cross section in the limit of high frequency of \\
scattered particle} \label{concl}

In the high-frequency limit $(k\rightarrow \infty)$ the first term
in Eq.(7) dies out, and the differential cross section takes form
\begin{eqnarray}
\frac{d\sigma}{d\varphi}&=&\frac{1}{2}r_c(1-4G\mu)^2\left|\sum\limits_{l}\sqrt{\cos[\frac{1}{2}
(1-4G\mu)(\varphi-\pi+2l\pi)]}\times \right.\nonumber \\
&\times& \left.\exp\{ie\Phi
l-2ikr_c\cos[\frac{1}{2}(1-4G\mu)(\varphi-\pi+2l\pi)]\}\right|^2,
\end{eqnarray}
where the summation is over integer $l$ satisfying condition
\begin{equation}
-\frac{\varphi}{2\pi}-\frac{2G\mu}{1-4G\mu}<l<-\frac{\varphi}{2\pi}+1+\frac{2G\mu}{1-4G\mu}.
\end{equation}
Note that results (10) and (11) are periodic in the value of flux
$\Phi$ with period equal to $2\pi e^{-1}$. This feature is common
with the scattering on a purely magnetic string ($\mu=0$). The
difference is that the Bohm-Aharonov differential cross section in
the low frequency limit ($k\rightarrow 0$) diverges in the forward
direction, $\varphi=0$, while Eq.(10) diverges in two symmetric
directions, $\varphi=\pm4G\mu(1-4G\mu)^{-1}$. The difference
becomes much more crucial in the high-frequency limit
$(k\rightarrow \infty)$. In the $\mu=0$ case one gets
\begin{equation}
\frac{d\sigma}{d\varphi}=\frac{1}{2}r_c\sin\frac{\varphi}{2},
\end{equation}
which is the cross section for scattering of a classical pointlike
particle by an impenetrable cylindrical shell of radius $r_c$;
evidently, the dependence on fractional part of $e\Phi(2\pi)^{-1}$
disappears in this limit. In the $\mu\neq 0$ case the dependence
survives, see Eq.(11). In particular, if $0<\mu<(8G)^{-1}$, which
is most interesting from the phenomenological point of view, then
the cross section at $k\rightarrow\infty$ takes the following form
in the region of the cosmic string shadow,
$-\frac{4G\mu\pi}{1-4G\mu}<\varphi<\frac{4G\mu\pi}{1-4G\mu}$:
\begin{eqnarray}
\frac{d\sigma}{d\varphi}&=&r_c(1-4G\mu)^2\left(\cos[\frac{1}{2}(1-4G\mu)\varphi]\sin(2G\mu\pi)
+\right.
\nonumber \\
&+&\left.\sqrt{\sin^2(2G\mu\pi)-\sin^2[\frac{1}{2}(1-4G\mu)\varphi]}\right.
\left.\cos\left\{e\Phi+4kr_c\sin[\frac{1}{2}(1-4G\mu)\varphi]\cos(2G\mu\pi)\right\}\right).
\end{eqnarray}
Integrating Eq.(14) over the region of the shadow and the
appropriate expression (which is independent of $\Phi$) over the
region out of the shadow, we obtain the total cross section in the
$k\rightarrow\infty$ limit:
\begin{equation}
\sigma_{\rm tot}=2r_c(1-4G\mu).
\end{equation}
The high-frequency limit is usually identified with the
quasiclassical limit. Although this identification is valid for
the total cross section, it is found to be invalid for the
differential cross section, see Eqs.(11) and (14) revealing the
periodic dependence on the flux, which is a purely quantum effect.

These results are generalized to the case of scattering of a
particle with spin.

\vskip 0.2cm

\section { Acknowledgements}

This work was supported by the State Foundation for Basic Research
of Ukraine (project 2.7/00152).

\vfill \eject

\end{document}